# Faster Approximation Schemes and Parameterized Algorithms on $H$-Minor-Free and Odd-Minor-Free Graphs

Siamak Tazari

Humboldt Universität zu Berlin
tazari@informatik.hu-berlin.de

**Abstract** We improve the running time of the general algorithmic technique known as Baker's approach (1994) on $H$-minor-free graphs from $\mathcal{O}(n^{f(|H|)})$ to $\mathcal{O}(f(|H|)n^{\mathcal{O}(1)})$ showing that it is fixed-parameter tractable w.r.t. the parameter $|H|$. The numerous applications include e.g. a 2-approximation for coloring and PTASes for various problems such as dominating set and max-cut, where we obtain similar improvements.

On classes of odd-minor-free graphs, which have gained significant attention in recent time, we obtain a similar acceleration for a variant of the structural decomposition theorem proved by Demaine et al. (2010) and a Baker-style decomposition into 2 graphs of bounded treewidth. We use these algorithms to derive faster 2-approximations; furthermore, we present the first PTASes and subexponential FPT-algorithms for independent set and vertex cover on these graph classes using a novel dynamic programming technique.

We also introduce a technique to derive (nearly) subexponential parameterized algorithms on $H$-minor-free graphs. We provide a *uniform* algorithm running in time $\inf_{0<\epsilon\leq 1}\mathcal{O}((1+\epsilon)^k + n^{\mathcal{O}(1/\epsilon)})$, where $n$ is the size of the input and $k$ is the number of vertices or edges in the solution. Our technique applies, in particular, to problems such as Steiner tree, (directed) subgraph with a property, (directed) longest path, and (connected/independent) dominating set, on some or all proper minor-closed graph classes, many of which were previously not even known to admit an algorithm with running time better than $\mathcal{O}(2^k n^{\mathcal{O}(1)})$. We obtain as a corollary that all problems with a minor-monotone subexponential kernel and amenable to our technique can be solved in subexponential FPT-time on $H$-minor free graphs. This results in a general methodology for subexponential parameterized algorithms outside the framework of bidimensionality.

**Keywords:** Parameterized Complexity; Graph Minors; Odd Minors; Subexponential Algorithms; Kernels.

## 1 Introduction

One of the most seminal results in algorithmic graph theory is arguably Baker's approach [1] for designing polynomial-time approximation schemes (PTAS) for a wide range of problems on planar graphs. Ever since its discovery, it has been applied and generalized in various ways, see e.g. [2,3,4,5,6,7,8,9]. The essence of the idea is the following: for any given $t$, one can partition a planar graph into $t$ parts, so that removing any one of the parts results in a graph of bounded treewidth. Now, to obtain a PTAS, we observe that if $t$ is appropriately chosen, there must exist a part that contains at most an $\epsilon$-fraction of an optimal solution; this can often be combined with the solution in the remainder of the graph to obtain a $(1 + \epsilon)$-approximation.

$H$-minor-free graphs, i.e. proper graph classes that are closed under building minors, have gained significant attention in the past two decades, especially due to Robertson and Seymour's graph minor theory, one of the deepest and most far-reaching theories in discrete mathematics in the past few decades. These classes include, e.g. planar graphs, bounded-genus graphs, linklessly embeddable graphs and apex graphs. Using the deep Robertson-Seymour (RS-) decomposition theorem [10], Grohe [4] generalized Baker's technique to $H$-minor-free graphs and Demaine et al. [7] showed the partitioning theorem mentioned above for all these graph classes. However, both their methods result in algorithms with running time $\mathcal{O}(n^{f(|H|)})$, for some computable function $f$; since $H$ is assumed to be fixed, this is considered polynomial.

**Improving Baker's Decomposition**. We provide the first algorithm for Baker's decomposition running in time $\mathcal{O}(g(|H|)n^{\mathcal{O}(1)})$, for some computable function $g$. This is a significant acceleration of the previous results, especially considering the fact that the constants in graph minor theory, such as the functions $f, g$ above, are usually huge. This immediately implies similar improvements on all the consequences of this algorithm, especially *all* the generic approximation algorithms and schemes in [4,7] and Baker's original problems [1]. In particular, we obtain the first 2-approximation for COLORING $H$-minor-free graphs in the given time bound and the first PTAS for

INDEPENDENT SET, MINIMUM COLOR SUM, MAX-CUT, MAXIMUM $P$-MATCHING, and DOMINATING SET on these graph classes while avoiding $|H|$ in the exponent of $n$ in their running time. Our main idea is derived from Dawar et al.'s approach [11] of finding a certain tree decomposition of $H$-minor-free graphs that is more tractable than the RS-decomposition.

**Parameterized Complexity**. In the language of parameterized complexity, our result above shows that partitioning $H$-minor-free graphs in the described way is *fixed-parameter tractable (*FPT*)* when parameterized by $|H|$. In this framework, for a given problem of size $n$ and parameter $k$, we are interested in algorithms with a running time of $\mathcal{O}(f(k)n^{\mathcal{O}(1)})$, where $f$ is some computable function depending solely on $k$; we then say that the problem at hand is in FPT. The theory also provides negative evidence that some problems are most likely not FPT; we refer to the books of Downey and Fellows [12] and Flum and Grohe [13] for more background on parameterized complexity.

Once a problem is shown to be FPT, the challenge is to provide algorithms that have the smallest dependence on the parameter $k$, i.e. make the function $f$ in the running time as small as possible. It is especially desirable to obtain *subexponential* functions and thus provide particularly fast algorithms. Whereas this is often not possible in general graphs, a plethora of results exist that show the existence of such algorithms on restricted graph classes, such as $H$-minor-free graphs. Perhaps the most general technique to obtain subexponential parameterized algorithms on these graph classes is the theory of *bidimensionality* [14] that captures almost all known results of this type on $H$-minor-free graphs. Still, this theory does not apply to a number of prominent problems, such as STEINER TREE, CONNECTED DOMINATING SET, and DIRECTED $k$-PATH.

One way to show fixed-parameter tractability of a problem is to show the existence of a *kernel* for the problem, which is a polynomial-time algorithm that transforms any instance into an equivalent instance whose size is bounded by a function $g$ solely dependent on $k$, called the *size* of the kernel. Kernelization can be seen as polynomial-time pre-processing with a quality guarantee and has gained much theoretical importance in the recent years – besides its natural practical importance. For an introduction to kernels we refer to the survey by Guo and Niedermeier [15].

**Guess and Conquer**. In this work, we provide a new framework, that we call *guess and conquer*, to obtain (nearly) subexponential parameterized algorithms on $H$-minor-free graphs for an abundant number of parameterized problems. Whenever the problem at hand admits a minor-monotone subexponential kernel, our method results indeed in a subexponential algorithm; otherwise, we obtain an algorithm with a running time $\mathcal{O}(2^{\mathcal{O}(\sqrt{k \log n})}n^{\mathcal{O}(1)}) = \inf_{0 < \epsilon \leq 1} \mathcal{O}((1+\epsilon)^k + n^{\mathcal{O}(1/\epsilon)})$ which we call *nearly* subexponential. Note that if $k = \mathcal{O}(\log n)$, our running time is fully polynomial in the input and if $k = \omega(\log n)$, it is subexponential FPT in $k$. Hence, except for a "small range" of possible parameter values, we have a subexponential FPT algorithm. In fact, we show that the problems we consider, admit a minor-monotone subexponential kernel on $H$-minor-free graphs if and only if they admit a subexponential FPT algorithm on these graph classes. Note that in general graphs, even a linear kernel results only in an exponential FPT-algorithm.

Our technique applies in particular to CONNECTED DOMINATING SET and STEINER TREE (at least) in bounded-genus graphs and DIRECTED $k$-PATH in all $H$-minor-free graphs, none of which are known to admit subexponential FPT-algorithms in $H$-minor-free graphs; for the latter two, such algorithms are not even known for planar graphs.

At the time of preparation of this paper, we became aware that Dorn et al. [16] recently and independently obtained similar nearly subexponential algorithms for some problems, albeit *only on apex-minor-free* graphs – whereas our techniques apply to general $H$-minor-free graphs. The focus of their work is on directed graph problems and in particular, they obtain a nearly subexponential FPT-algorithm for DIRECTED $k$-PATH in apex-minor-free graphs (furthermore, they obtain a number of subexponential FPT-algorithms for problems that we do not consider in this work). The second method they present in their paper is indeed similar to what we call guess and conquer in this work but it is only formulated for a specific problem, not in general terms as we do, and also not for the wide range of problems we consider; in particular, our technique for domination and covering problems is completely new. Additionally, their method works only for problems with a polynomial kernel whereas we also obtain nearly subexponential algorithms on problems without such a kernel. Furthermore, even on problems with a kernel we obtain faster algorithms, having a running time of $\mathcal{O}(2^{\mathcal{O}(\sqrt{k \log k})}n^{\mathcal{O}(1)})$ instead of $\mathcal{O}(2^{\mathcal{O}(\sqrt{k} \log k)}n^{\mathcal{O}(1)})$ of [16].

**Odd-Minor-Free Graphs**. The class of odd-minor-free graphs has attained extensive attention in the graph theory literature [17,18] and recently, in theoretical computer science [19,20,21]. They are strictly more general than $H$-minor-free graphs as they include, for example, all bipartite graphs and may contain a quadratic number of edges. In addition to their role in graph minor theory and structural graph theory, they bear important connections to the MAX-CUT problem [17] and Hadwiger's conjecture [22,19]. We refer to the work of Demaine et al. [20] for a more thorough introduction to odd-minor-free graphs and their significance.

Demaine et al. [20] prove a decomposition theorem for odd-$H$-minor-free graphs that is similar to the RS-decomposition of $H$-minor-free graphs and present an $\mathcal{O}(n^{f(|H|)})$ algorithm to compute such a decomposition. From this, they derive a Baker-style decomposition of odd-minor-free graphs into 2 graphs of bounded treewidth. We identify an intermediate decomposition implicit in [20] that is computable in FPT-time and proves to be very useful algorithmically: on one hand, we deduce the Baker-style decomposition into 2 parts and a number of 2-approximation algorithms (most notably for COLORING) in FPT-time as a corollary; on the other hand, we can answer a question that is posed several times by Demaine et al. in [20], affirmatively: nameley, whether the PTASes and subexponential FPT-algorithms for VERTEX COVER and INDEPENDENT SET can be generalized from $H$-minor-free graphs to odd-minor-free graphs. We show how to obtain such algorithms by introducing a novel dynamic programming technique on odd-minor-free graphs based on solving a certain weighted version of the considered problems in bipartite graphs. These are the first PTASes and subexponential FPT-algorithms developed on odd-minor-free graphs.

**Organization**. We recall some relevant notions from graph theory and parameterized complexity in Section 2. The partitioning algorithm and its implications are presented in Section 3. In Section 4, we state and discuss our technique for (nearly) subexponential FPT-algorithms. Our algorithms on odd-minor-free graphs are presented in Section 5 and we conclude in Section 6.

## 2 Preliminaries

We collect the relevant concepts from parameterized complexity and graph theory.

**Subexponential Algorithms and FPT**. We use the standard notions of parameterized complexity and bounded fixed-parameter tractability [13]. A parameterized problem is said to be *fixed-parameter tractable* (FPT) if for any instance of size $n$ with parameter $k$ it can be solved in time $f(k)n^{\mathcal{O}(1)}$, for some computable function $f$ solely dependant on $k$. If $f$ belongs to a class $\mathcal{F}$ of functions from $\mathbb{N}$ to $\mathbb{N}$, we say that our problem is in $\mathcal{F}$-FPT. We denote $2^{k^{\mathcal{O}(1)}}$-FPT, $2^{\mathcal{O}(k)}$-FPT, and $2^{o^{\text{eff}}(k)}$-FPT by EXPT, EPT, and SUBEPT, respectively; here, $f \in o^{\text{eff}}(g)$ if there exists $n_0 \in \mathbb{N}$ and a computable, nondecreasing and unbounded function $\iota : \mathbb{N} \to \mathbb{N}$, such that $f(n) \leq \frac{g(n)}{\iota(n)}$ for all $n \geq n_0$. A problem is *subexponential fixed-parameter tractable* if it is in SUBEPT.

Observe that if a problem is in SUBEPT then there exists an algorithm for the problem, so that for any fixed $\alpha > 0$ the algorithm runs in time $\mathcal{O}(2^{\alpha k} n^{\mathcal{O}(1)})$. We define a problem to be in SUBEPT$^+$ if it is solved by an algorithm $\mathcal{A}$, so that for any fixed $\alpha > 0$, the running time of $\mathcal{A}$ is $\mathcal{O}(2^{\alpha k} n^{\mathcal{O}(1/\alpha)})$. Observe that we require a *single* (uniform) algorithm to have this property for the considered problem. Clearly, SUBEPT $\subseteq$ SUBEPT$^+$ $\subseteq$ EPT. Note that the *non-uniform exponential time hypothesis (ETH)* implies that SUBEPT$^+ \neq$ EPT.

**Kernels**. A *kernel* of a parameterized problem is a fully polynomial algorithm that given an instance of size $n$ and parameter $k$, returns a *equivalent reduced instance* of the same problem of size $f(k)$ and parameter $k' \leq k$. The function $f$ denotes the *size* of the kernel; we then speak of a *linear, polynomial, and subexponential* kernel, respectively. Kernelization is a major technique in fixed parameter complexity as any computable parameterized problem is in FPT if and only if it admits a kernel [23].

**Graphs and Minors**. We usually denote graphs by letters $G, H$, and refer to their vertex/edge sets by $V(G)$ and $E(G)$, respectively. Unless otherwise mentioned, our graphs have $n$ vertices and $m$ edges. For a subset $U \subseteq V(G)$, we write $G[U]$ to denote the subgraph of $G$ induced by $U$. The *r-neighborhood* of a vertex $v$, denoted by $N_r(v)$, is the set of vertices at distance at most $r$ from $v$; we define $N(v) = N_1(v)$. We let $d_G(u,v)$ denote the distance between vertices $u, v \in V(G)$. For an edge $e = uv$ in $G$, we define the operation $G/e$ of *contracting* $e$ as identifying $u$ and $v$ and removing all loops and duplicate edges. A graph $H$ is a *minor* of $G$, written as $H \preceq G$, if it can be obtained from $G$ by a series of vertex and edge deletions and contractions. We say $G$ is an *$H$-minor-free graph* if it does not contain $H$ as a minor. A class of graphs that is closed under building minors and does not contain all graphs is called a *proper minor-closed class of graphs*. A class of graphs is a proper minor-closed class if and only if it is $H$-minor-free for some fixed $H$. Examples of such classes include planar graphs, bounded-genus graphs, and linklessly embeddable graphs. An *apex-graph* is a planar graph augmented by an additional vertex that can have edges to any other vertex. A class of graphs is called *apex-minor-free* if it excludes a fixed apex-graph as a minor. It is a well-known fact that $H$-minor-free graphs have bounded average degree (depending only on $|H|$), i.e. they fulfill $m = \mathcal{O}_H(n)$; we use the notation $\mathcal{O}_H$ to denote that the constants hidden in the big-$\mathcal{O}$ depend on $|H|$.[1] We refer the reader to [24] for more background on graph theory.

---

[1] this is necessary since in graph minor theory, the exact dependence is often not known.

**Odd Minors**. A *model* of $H$ in $G$ is a map that assigns to every vertex $v$ of $H$, a connected subtree $T_v$ of $G$ such that the images of the vertices of $H$ are all disjoint in $G$ and there is an edge between them if there is an edge between the corresponding vertices in $H$. A graph $H$ is a minor of $G$ if and only if $G$ contains a model of $H$. Now $H$ is an *odd-minor* of $G$ if additionally the vertices of the trees in the model of $H$ in $G$ can be 2-colored in such a way that (i) the edges of each tree $T_v$ are bichormatic; and (ii) every edge $e_G$ in $G$ that connects two trees $T_u$ and $T_v$ and corresponds to an edge $e_H = uv$ of $H$ is monochromatic. A graph is *odd-$H$-minor-free* if it excludes $H$ as an odd minor. For example, bipartite graphs are odd-$K_3$-minor-free.

**Tree Decompositions and Dynamic Programming**. A *tree decomposition* of a graph $G$ is a pair $(T, \mathcal{B})$, where $T$ is a tree and $\mathcal{B} = \{B_u | u \in V(T)\}$ is a family of subsets of $V(G)$, called *bags*, such that (i) every vertex of $G$ appears in some bag of $\mathcal{B}$; (ii) for every edge $e = \{u, v\}$ of $G$, there is a bag of $\mathcal{B}$ containing $\{u, v\}$; and (iii) for every vertex $v \in V(G)$ the set of bags containing $v$ forms a connected subtree $T_v$ of $T$. The *width* of a tree decomposition is the maximum size of any bag in $\mathcal{B}$ minus 1. The *treewidth* of $G$, denoted by $\text{tw}(G)$, is the minimum width over all possible tree decompositions of $G$. The *adhesion* of a tree decomposition is defined as $\max\{|B_u \cap B_t| \mid \{u, t\} \in E_T\}$. Many NP-hard optimization problems become fixed-parameter tractable when parameterized by the treewidth of the instance, by using dynamic programming on a given tree decomposition. The most well-known result in this area is Courcelle's theorem [25] stating that any problem definable in monadic second-order logic is in FPT when parameterized by the treewidth and the length of the formula. However, the algorithms obtained by this theorem usually have multiply-exponential dependence on the treewidth of $G$. In this work, we are interested in algorithms with singly-exponential dependence on the treewidth, i.e. problems that are in EPT when paramterized by treewidth. Several natural problems have long been known to admit such algorithms [26,27], and for problems with global connectivity requirement such as LONGEST PATH, STEINER TREE and CONNECTED DOMINATING SET, Dorn et al. [28] recently gave EPT-algorithms on (some classes of) $H$-minor-free graphs by utilizing Catalan structures.

**On Local Treewidth**. We say that a graph has *bounded local treewidth* if for every vertex $v$ and integer $r$, we have $\text{tw}(N_r(v)) \leq f(r)$, for some computable function $f$ depending solely on $r$; we write $\text{ltw}_r(G) \leq f(r)$. Demaine and Hajiaghayi [29] showed that every minor-closed class of graphs that has bounded local treewidth has, in fact, *linear* local treewidth, i.e. in this case, we have $\text{ltw}_r(G) \leq \lambda r$ for a fixed integer $\lambda$ depending only on the excluded minor of the class. Eppstein [2] showed that a minor-closed class of graphs has bounded local treewidth if and only if it is apex-minor-free.

## 3 Partitioning $H$-minor-free graphs

In [7], Demaine et al. show how to decompose $H$-minor free graphs into parts, so that upon removal of any part, the problem at hand becomes tractable. In this section, we show how this decomposition can be achieved in FPT-time with $|H|$ as parameter; furthermore, we introduce a refinement of this method.

**Theorem 1 (Demaine et al. [7]).** *For every graph $H$ there is a constant $c_H$ such that for any integer $p \geq 1$ and for every $H$-minor-free graph $G$, the vertices (edges) of $G$ can be partitioned into $p$ sets such that any $p - 1$ of the sets induce a graph of treewidth at most $c_H p$. Furthermore, such a partition can be found in polynomial time.*

The essence of this idea goes back to Baker's approach [1] for polynomial-time approximation schemes on planar graphs. That approach has been applied and generalized in many ways [2,3,4,5,6,7,8,9]. By now, it is considered a standard technique and it is not hard to see that it is true for apex-minor-free graphs: we perform breadth-first search (BFS) and label the BFS-layers periodically by $0, \ldots, p-1$; now by removing the vertices of any one label, the graph falls apart into a number of connected components. If we consider such a connected component $C$ in the BFS-tree then, by *contracting* everything preceding $C$ into a single vertex and *deleting* everything following $C$ in the tree, we obtain a graph of bounded diameter. Such a graph has bounded treewidth because of the bounded local treewidth property of apex-minor-free graphs, and so we obtain a linear time algorithm.

But for general $H$-minor-free graphs the situation is more complicated; the bounded local treewidth property does not hold for all $H$-minor free graphs. To overcome this difficulty, Demaine et al. [7] apply the *Robertson-Seymour decomposition (RS-decomposition)* of $H$-minor-free graphs [10], following some ideas of Grohe [4]. They give an algorithm to compute an RS-decomposition of a given $H$-minor-free graph in time $n^{\mathcal{O}_H(1)}$. For parameter $|H|$, that algorithm is hence not a fixed-parameter algorithm. We now establish fixed-parameter versions and extensions of Theorem 1.

### 3.1 The Existence of a Fast Partitioning Algorithm

A key observation to obtain an FPT-version of the partitioning algorithm is that an RS-decomposition is not needed – it suffices to have a tree decomposition of the input graph that fulfills certain properties. To state these properties, we require the following classes of graphs as defined by Grohe [4]:

$$\mathcal{L}(\lambda) = \{G \mid \forall H \preceq G \, \forall r \geq 1 \, : \, \text{ltw}_r(H) \leq \lambda \cdot r\},$$
$$\mathcal{L}(\lambda, \mu) = \{G \mid \exists U \subseteq V(G) \, : \, |U| \leq \mu \text{ and } G - U \in \mathcal{L}(\lambda)\} \, .$$

Since the property of having bounded local treewidth is not inherited when taking minors, we explicitly require it for all minors of $G$ in the definition of $\mathcal{L}(\lambda)$. A graph $G$ in the class $\mathcal{L}(\lambda, \mu)$ may contain a set $U$ of at most $\mu$ *apices*, so that by removing these apices from $G$ we obtain a graph in $\mathcal{L}(\lambda)$. Note that both of these classes are minor-closed and hence, by the Graph Minor Theorem and the minor-testing algorithm of Robertson and Seymour [30,31], they can be recognized in time $\mathcal{O}_H(n^3)$.

Given a graph $G$, consider a tree decomposition $(T, \mathcal{B})$ of $G$ and a bag $B$ in $\mathcal{B}$. The *closure* of $B$, denoted by $\overline{B}$, is the graph obtained from $G[B]$ by adding some edges so that $B \cap B'$ induces a clique in $\overline{B}$, for every bag $B' \neq B$. We say $G$ has a tree decomposition *(strongly) over* a class of graphs $\mathcal{C}$ if there exists a tree decomposition of $G$ so that the closure of each bag is in $\mathcal{C}$. Note that if $\mathcal{C}$ is minor-closed then the class of graphs having a tree decomposition over $\mathcal{C}$ is minor-closed, too. Using the Robertson-Seymour decomposition theorem [10], Grohe [4] proved that for every $H$ there exist computable $\lambda$, $\mu$, and $\kappa$ depending only on $|H|$, so that any $H$-minor-free graph admits a tree decomposition over $\mathcal{L}(\lambda, \mu)$ with *adhesion at most* $\kappa$. Later [32] he observed that such a tree decomposition can be computed in time $\mathcal{O}_H(n^5)$, provided that the excluded minors of the class of graphs having such a decomposition are known. In fact, he proved the *existence* of such an algorithm for all proper minor-closed graph classes without presenting it *explicitly* for any particular class; a common fate when applying the Graph Minor Theorem. Furthermore, this algorithm is *non-uniform* in the sense that for every excluded minor $H$, we obtain a different algorithm. Nevertheless, based on that decomposition, the proof of Theorem 1 can easily be adapted to obtain

**Theorem 2.** *There exists an algorithm that computes a partition as described in Theorem 1 in time $\mathcal{O}_H(n^5)$.*

### 3.2 An Explicit FPT-algorithm

The statement of Theorem 2 is not quite satisfactory; we would like to *know* and furthermore, have a *uniform* algorithm to compute the desired decomposition. Dawar et al. [11] attacked this problem in the following way: instead of looking at the *closure* of bags in a tree decomposition, they look at the *companion* of the bags: for a bag $B$ with neighbors $B_1, \ldots, B_t$ in a given tree decomposition, we define its companion $\hat{B}$ as the graph obtained from $G[B]$ by adding new vertices $\hat{u}_1, \ldots, \hat{u}_t$ and connecting $\hat{u}_i$ to all vertices in the intersection $B \cap B_i$, for $1 \leq i \leq t$. We call $\hat{u}_1, \ldots, \hat{u}_t$ the *hat vertices of* $\hat{B}$. Note that the difference between the closure and the companion of a bag is that in the closure, the intersections with neighboring bags form a clique instead of being connected to a hat vertex. We say that a graph $G$ has a tree decomposition *weakly over* a graph class $\mathcal{C}$ if there exists a tree decomposition of $G$ so that the companions of all the bags are in $\mathcal{C}$.

**Theorem 3 (Dawar et al. [11]).** *There is an explicit uniform algorithm that, given an $H$-minor-free graph $G$, computes a tree decomposition $(T, \mathcal{B})$ of $G$ that is weakly over $\mathcal{L}(\lambda, \mu)$ and has adhesion at most $\kappa$, in time $\mathcal{O}_H(n^{\mathcal{O}(1)})$, where $\lambda$, $\mu$, and $\kappa$ are computable functions depending only on $|H|$. Furthermore, the $\mu$ apices of the companion of each bag in $\mathcal{B}$ can be computed in the same time bound.*

Note that the $\lambda$, $\mu$, and $\kappa$ in the theorem above are much larger than the ones in the existential version proven by Grohe [4]; but they still depend solely on $|H|$ and are thus acceptable for our purposes. However, in order to provide an explicit FPT-algorithm for Theorem 1, we actually need the closure of the bags of the tree decomposition to be in $\mathcal{L}(\lambda, \mu)$. We resolve this issue by using Lemma 5 below. First, we need some preparation:

Let $G$ be a graph and let $(T, \mathcal{B})$ be a tree decomposition of $G$ with adhesion at most $\kappa$ that is weakly over $\mathcal{L}(\lambda, \mu)$, for some $\kappa$, $\lambda$, and $\mu$, and assume $T$ is rooted at some bag. We say the apex set $A$ of the companion $\hat{B}$ of a bag $B \in \mathcal{B}$ is *nice*, if (i) for the parent $B'$ of $B$ in $T$, we have $B \cap B' \subseteq A$; and (ii) if $\hat{u}$ is a hat vertex of $\hat{B}$ belonging to $A$, then $N(\hat{u}) \subseteq A$. Note that by going from $\mathcal{L}(\lambda, \mu)$ to $\mathcal{L}(\lambda, \mu\kappa + \kappa)$ if necessary, we may assume w.l.o.g that all the apex sets of the companions of the bags are nice: simply add the intersection with the parent bag and all the neighbors of included hat vertices to each apex set. We proceed with our main technical lemmas[2].

---

[2] Lemma 4 is implicitly assumed by Dawar et al. [11]

**Lemma 4.** *Let $G$ be an $H$-minor-free graph and let $(T, \mathcal{B})$ be a tree decomposition of $G$ with adhesion at most $\kappa$ that is weakly over $\mathcal{L}(\lambda, \mu)$. Consider a bag $B_0 \in \mathcal{B}$ with nice apex set $A \subseteq V(\hat{B}_0)$ and closure $\overline{B}_0$. Define $B := B_0 - A$ and $\overline{B} := \overline{B}_0 - A$. Let $j \geq i \geq 0$ be integers, $r \in B$, and $L_{i,j}^r := \{v \in B \mid i \leq d_{\overline{B}}(r, v) \leq j\}$. Then we have $\mathrm{tw}(\overline{B}[L_{ij}^r]) \leq 4\lambda\kappa \cdot (j - i + 1)$.*

*Proof.* Let $\hat{U}$ be the set of hat vertices of $\hat{B}_0$ and $\hat{B} := \hat{B}_0 - A$. For $v \in B$, we let $d(v)$ denote $d_{\overline{B}}(r, v)$ and define $p := j - i + 1$. For a subgraph $\overline{C}$ of $\overline{B}$, we define $C := V(\overline{C})$ and its companion $\hat{C}$ as the subgraph of $\hat{B}$ induced by $C$ together with those hat vertices in $\hat{U}$ that have a neighbor in $C$. Note that $\overline{C}$ is connected if and only if $\hat{C}$ is connected and that the diameter of $\hat{C}$ is at most twice that of $\overline{C}$; furthermore, $\hat{C}$ remains disjoint from $A$ because $A$ is nice.

From now on, fix $\overline{C} := \overline{B}[L_{ij}^r]$. Let $R$ be the set of all $v \in B$ with $d(v) < i$; let $\overline{R} := \overline{B}[R]$ and $\hat{R}$ its companion. Since $\overline{R}$ is connected, $\hat{R}$ is connected, too; but there might exist a set $\hat{U}' := V(\hat{C}) \cap V(\hat{R})$ of hat vertices that are contained in both $\hat{R}$ and $\hat{C}$. We claim that $\hat{R}' := \hat{R} - \hat{U}'$ is still connected; to see this, let $\hat{u} \in \hat{U}'$ be such a hat vertex and note that $\hat{u}$ has only neighbors $N_R \subseteq R$ and $N_C \subseteq C$, so that $N_R \cup N_C$ induces a clique in $\overline{B}$. But then, it must be that the vertices $v_R \in N_R$ fulfill $d(v_R) = i - 1$ and the ones $v_C \in N_C$ fulfill $d(v_C) = i$ and so, the vertices of $N_R$ are connected to $r$ via a path that does not include any of the edges of this clique. Hence $\hat{R} - \hat{u}$ is connected.

Now let $Q := C \cup R$, $\overline{Q} := \overline{B}[Q]$, and $\hat{Q}$ its companion. Consider the graph $\hat{Q}' := \hat{Q}/E(\hat{R}')$ obtained by contracting $\hat{R}'$ in $\hat{Q}$. Since $\hat{R}'$ is connected and disjoint from $\hat{C}$, we observe that $\hat{Q}'$ is isomorphic to $\hat{C}$ augmented by a single vertex $r'$ that is connected to all vertices of $v \in C$ with $d(v) = i$ – either by a direct edge or by using a hat vertex from $\hat{U}'$. Hence, the distance of any vertex $v \in C$ from $r'$ is at most $2p$ and the diameter of $\hat{Q}'$ is at most $4p$. On the other hand, we have $\hat{Q}' \preceq \hat{B} \in \mathcal{L}(\lambda)$ and so, the treewidth of $\hat{Q}'$ is bounded by $4\lambda \cdot p$.

Let $(T', \mathcal{B}')$ be a tree decomposition obtained by (i) considering a tree decomposition $(T_0, \mathcal{B}_0)$ of $\hat{Q}'$ of width at most $4\lambda \cdot p$; (ii) removing the vertex $r'$ from every bag in $\mathcal{B}_0$; and (iii) replacing every hat vertex in each bag in $\mathcal{B}_0$ by the set of its neighbors. Clearly, $(T', \mathcal{B}')$ is a tree decomposition of $\overline{C}$ and its width is at most $4\lambda\kappa \cdot p$, as desired. □

**Lemma 5.** *Let $G$ be an $H$-minor-free graph and let $(T, \mathcal{B})$ be a tree decomposition of $G$ with adhesion at most $\kappa$ that is weakly over $\mathcal{L}(\lambda, \mu)$, where $\lambda$, $\mu$, and $\kappa$ are computable functions depending only on $|H|$. Consider a bag $B_0 \in \mathcal{B}$ with a given nice apex set $A \subseteq V(\hat{B}_0)$. For any integer $p \geq 1$ we can label the vertices and edges of the closure $\overline{B}_0 - A$ by the numbers $\{0, \ldots, p - 1\}$, so that the following holds:*

1. *every edge has the label of one of its endpoints;*
2. *every vertex is incident to at most 2 distinct edge-labels;*
3. *the vertices or edges of any $p - 1$ labels induce a graph of treewidth at most $4\lambda\kappa \cdot p$;*
4. *there exists an explicit uniform algorithm that finds such a labeling in time $\mathcal{O}_H(n^{\mathcal{O}(1)})$.*

*Proof.* Define $B := B_0 - A$ and $\overline{B} := \overline{B}_0 - A$. W.l.o.g. we may assume $\overline{B}$ is connected, since otherwise we can simply repeat the following procedure for every connected component. We pick an arbitrary vertex $r$ and perform a BFS in $\overline{B}$. We assign the label $\mathrm{lab}(v) = d_{\overline{B}}(v, r) \mod p$ to every vertex $v \in V$; every edge is assigned the label of its endpoint that is closer to $r$. Consider a connected component $\overline{C}$ of the graph induced by any $p - 1$ vertex or edge labels. Then $\overline{C}$ is a subgraph of $L_{i,i+p-1}^r$, for some $i \geq 0$, and hence, by Lemma 4 its treewidth is bounded by $4\lambda\kappa \cdot p$. The other claims are easy to verify. □

Using Theorem 3 and Lemma 5, it is not hard to extend the proof of Theorem 1 in [7] to obtain the following version; on the other hand, the proof of Theorem 7 as given below also implies this result:

**Theorem 6.** *There exists an explicit uniform algorithm that computes a partition as described in Theorem 1 and runs in time $\mathcal{O}_H(n^{\mathcal{O}(1)})$.*

### 3.3 Bounding the Number of Label Incidences

In some of our applications, we need a more specific version of Theorem 1; we would like to obtain a partition of the *edges* while still being able to bound the number of parts in which each *vertex* might appear. To this end, we shall bound the number of distinct edge-labels incident to each vertex in an edge-partition of the graph. A closer look at Demaine et al.'s [7] proof of Theorem 1 reveals that this number is indeed bounded by $\mathcal{O}_H(1)$; for the sake of completeness, and since our setting is somewhat different, we include a proof in this section.

For two graphs $G_1$ and $G_2$ whose intersection $E(G_1) \cap E(G_2)$ induces a clique, we define their *clique-sum* $G_1 \oplus G_2$ as the graph $G_1 \cup G_2$ with any number of edges in the clique $E(G_1) \cap E(G_2)$ deleted. Note that this operation is not well-defined and can have a number of possible outcomes. The notion of a clique-sum plays a central role in graph minor theory and it is well-known that $\mathrm{tw}(G_1 \oplus G_2) \leq \max\{\mathrm{tw}(G_1), \mathrm{tw}(G_2)\}$. A key observation is that when considering two neighboring bags $B_1$ and $B_2$ in a tree decomposition of a graph, the graph induced by $B_1 \cup B_2$ is a subgraph of a clique-sum $\overline{B_1} \oplus \overline{B_2}$ of the closure of the bags. We use this observation to prove the following theorem.

**Theorem 7.** *For any fixed graph $H$ there are constants $c_H$ and $d_H$ such that for any integer $p \geq 1$ and every $H$-minor-free graph $G$, the edges of $G$ can be partitioned into $p$ parts such that any $p-1$ of the parts induce a graph of treewidth at most $c_H p$ and every vertex appears in at most $d_H$ of the parts. Furthermore, such a partition can be found in explicit uniform* FPT*-time, i.e. $\mathcal{O}_H(n^{\mathcal{O}(1)})$.*

*Proof.* First, we compute a tree decomposition $(T, \mathcal{B})$, weakly over $\mathcal{L}(\lambda, \mu)$ with adhesion $\kappa$ as given by Theorem 3. We root the tree decomposition at a bag $B_0$ and let $B_0, \ldots, B_k$ be a pre-order traversal of the bags of the rooted tree decomposition. For $0 \leq i \leq k$, let $G_i := \overline{B}_0 \oplus \cdots \oplus \overline{B}_i$; note that $G = G_k$. For each $i \in \{1, \ldots, k\}$, let $C_i$ be the set of at most $\kappa$ vertices in the intersection of $B_i$ with its parent bag and $C_0 = \emptyset$; also, let $\hat{A}_i$ be the set of at most $\mu$ apex vertices of the companion $\hat{B}_i$ and assume w.l.o.g. that $\hat{A}_i$ is nice, i.e. in particular, $C_i \subseteq \hat{A}_i$; let $A_i = \hat{A}_i \cap B_i$. We prove the statement by induction on $i$, label the vertices and edges simultaneously, and keep the invariant that the label of every edge is equal to the label of one of its endpoints.

For $G_0 = \overline{B}_0$, we start with the labeling provided by Lemma 5. Next, we assigning the label 0 to all vertices and edges that are included or have an endpoint in $A_0$. Since Lemma 5 guarantees that every vertex in $B_0 - A_0$ is incident to at most 2 distinct edge-labels, we obtain that in $G_0$, every vertex is incident to at most 3 distinct edge-labels. Also, the treewidth of any subgraph of $G_0$ induced by any $p-1$ labels is at most $c_H p$ with $c_H := 4\lambda\kappa + \mu$, as one can simply add all the vertices of $A_0$ to every bag in a tree decomposition provided by Lemma 5.

Now consider $G_i$ for $1 \leq i \leq k$ and let $B_j$ be the parent bag of $B_i$. We consider the labeling inductively constructed for $G_{i-1}$ and label the vertices and edges of $\overline{B}_i - A_i$ using Lemma 5. Since $C_i = B_i \cap B_j = B_i \cap V(G_{i-1}) \subseteq A_i$, we let $C_i$ and all the edges with both endpoints in $C_i$ keep the labels obtained in $G_{i-1}$ without causing a conflict; we label the vertices in $A_i - C_i$ by 0, each edge with exactly one endpoint $u$ in $C_i$ by the label of $u$ and all remaining edges (with an endpoint in $A_i$) by 0. Note that the number of incident edge-labels for each vertex in $G_{i-1}$ does not change; and every vertex in $B_i - C_i$ becomes incident to at most $d_H := \kappa + 2$ different labels. Hence, this requirement of the theorem is fulfilled by induction.

It remains to show that any subgraph $G'_i$ of $G_i$ induced by (at most) $p-1$ labels has treewidth at most $c_H p$. Let $G'_{i-1} := G'_i[V(G_{i-1}) \cap V(G'_i)]$ and $\overline{B}'_i := G'_i[B_i]$. Let $d$ be the omitted label and $D \subseteq C_i$ the set of vertices with label $d$ from $C_i$; define $C' := C - D$ and $\overline{B}''_i := \overline{B}'_i - D$. Note that $C'$ induces a clique in both $G'_{i-1}$ and $\overline{B}''_i$ and that $D$ has no neighbors in $B_i - C_i$ in $\overline{B}'_i$ because all edges incident to a vertex of $D$ in $\overline{B}'_i$ have label $d$ by our construction and are deleted. Hence, $G'_i = G'_{i-1} \oplus \overline{B}''_i$ and so its treewidth is bounded by the maximum of the treewidth of these two graphs. But by the induction hypothesis and Lemma 5, this number is bounded by $c_H p$. □

### 3.4 Approximation Algorithms and PTAS

We improve *all* the generic approximation and PTAS results given by Demaine et al. in [7] (specifically, Theorems 3.3–3.7) and also by Grohe in [4] by removing the dependence on $|H|$ from the exponent of $n$ in the presented algorithms. This is due to Theorem 6 and also the fact that Lemma 4 corresponds to Lemma 16 in [4] as applied to tree decompositions that are *weakly* over $\mathcal{L}(\lambda, \mu)$. Nothing else in the proofs and algorithms needs to be changed. We refrain from re-stating all the generic results and highlight only some important concrete corollaries below.

**Corollary 8.** *There exists a 2-approximation algorithm for* COLORING *an $H$-minor-free graph in time $\mathcal{O}_H(n^{\mathcal{O}(1)})$.*

**Corollary 9.** *There exists a PTAS for* INDEPENDENT SET, VERTEX COVER, MINIMUM COLOR SUM, MAX-CUT, *and* MAXIMUM $P$-MATCHING *in $H$-minor-free graphs running in time $\mathcal{O}_{H,\epsilon}(n^{\mathcal{O}(1)})$.*

The following result is of particular interest, as it does not follow from Theorem 6 but requires the techniques of [4]:

**Corollary 10.** *There exists a PTAS for* DOMINATING SET *in $H$-minor-free graphs running in time $\mathcal{O}_{H,\epsilon}(n^{\mathcal{O}(1)})$.*

For all the problems mentioned above, our method results in the first algorithm with this running time. The class of problems to which these techniques apply is very large and includes all the problems originally considered by Baker [1] and also most minor-bidimensional problems, whereas for the latter, other known techniques also result in such PTASes [33,34].

# 4 Guess and Conquer for Subexponential FPT-Algorithms

In this section, we introduce the technique of *guess and conquer* that for a wide range of problems shows their membership in SUBEPT$^+$. We first present the technique for the generic problem of finding a subgraph with a property; then we show how the idea can be used for domination, covering, and further types of problems. At the end of the section, we discuss the relation of the proposed algorithm to polynomial and subexponential kernels.

## 4.1 The Technique

We state our main technique for a broad class of parameterized problems. Given a *graph property* $\pi$, which is simply a set of directed or undirected graphs, we consider the following generic problem:

> $k$-SUBGRAPH WITH PROPERTY $\pi$: Given a graph $G$, does $G$ contain a subgraph with at most $k$ vertices that has property $\pi$, i.e. is isomorphic to some graph in $\pi$?

The problem is abbreviated as $k$-SP($\pi$). If we insist on finding *induced* subgraphs with property $\pi$, we use the notation $k$-ISP($\pi$) and if we want $k$ to be the number of edges in an edge-induced subgraph then the problem is denoted by $k$-EISP($\pi$). We allow that some vertices in the graphs in $\pi$ have *fixed labels*, in which case, the task becomes to find a subgraph of a (partially) labeled graph $G$ isomorphic to a graph in $\pi$, so that the labels match. Another variant is that we are additionally given a set $R \subseteq V(G)$ of *roots* (or terminals) in $G$ and we are seeking a subgraph with property $\pi$ that contains all the roots. We use the letters $L$ and $R$ to account for the labeled and rooted version of the problem, respectively, and the letter $D$ to emphasize that we are dealing with directed graphs. Finally, we might be given a vertex- or edge-weighted graph and our goal is to find among all subgraphs of $G$ with property $\pi$ and at most $k$ vertices (or edges), the one of *minimum* or *maximum* weight. We denote this whole class of problems by $(\{\text{MIN},\text{MAX}\})\ k\text{-}\{D,R,L,E,I\}\text{SP}(\pi)$.

For example, the parameterized version of the STEINER TREE problem can be seen as a MIN $k$-RSP($\pi$) problem, where $\pi$ is the set of all trees and $R$ is the set of terminals that are to be connected in $G$. Likewise, one could look for a biconnected subgraph of size at most $k$ containing a given set of terminals by taking $\pi$ to be the set of all biconnected graphs. Also, DIRECTED $st$-$k$-PATH is an instance of $k$-DLEISP($\pi$) where $\pi$ contains only a directed path of length $k$, in which the first vertex is labeled $s$ and the last vertex is labeled $t$. Other interesting choices for $\pi$ include being chordal, bipartite, edge-less (INDEPENDENT SET), of maximum degree $r \geq 1$, a clique, planar, or containing only/avoiding cycles of specified length [7].

We obtain the following general result:

**Theorem 11.** *Let $\pi$ be a graph property such that on graphs of treewidth $t$ one can find a (maximum/minimum weight/rooted/labeled/induced) subgraph with property $\pi$ in time $\mathcal{O}(2^{\mathcal{O}(t)}n^{\mathcal{O}(1)})$. For any (directed/partially labeled) graph $G$ excluding a fixed graph $H$ as a minor, there exists an algorithm $\mathcal{A}$ solving problem $(\{\text{MIN},\text{MAX}\})\ k\text{-}\{D,R,L,E,I\}\text{SP}(\pi)$ and that for any $\alpha \geq 1$, $\epsilon \in (0,1]$ and fixed $\delta > 0$ runs in time $\mathcal{O}_H(2^{\mathcal{O}_H(\sqrt{k \log n})}n^{\mathcal{O}(1)}) = \mathcal{O}_H(2^{\mathcal{O}_H(k/\alpha)} + n^{\mathcal{O}(\alpha)}) = \mathcal{O}_H((1+\epsilon)^k + n^{\mathcal{O}_H(1/\epsilon)}) = o(n^{\mathcal{O}(1)+\delta\sqrt{k}})$. In particular, the considered problem belongs to SUBEPT$^+$.*

*Proof.* Let $p$ be some fixed integer; apply Theorem 6 to $G$ to obtain a partition $V_1, \ldots, V_p$ of the vertex set of the graph, so that the graph induced by any $p-1$ of the sets has treewidth at most $c_H p$; such partition can be found in time $\mathcal{O}_H(n^{\mathcal{O}(1)})$. Now, consider an optimal subgraph $S^\star$ fulfilling the requirements of the problem; since $S^\star$ is assumed to have at most $k$ vertices, there exists an $i^\star \in \{1, \ldots, p\}$, so that $V_{i^\star}$ contains at most $\lfloor k/p \rfloor$ vertices of $S^\star$. Since we do not know the value of $i^\star$, we simply *guess* it; there are at most $p$ possibilities to do so and we try all of them. Hence, for each $i \in \{1, \ldots, p\}$, we repeat the following:

For a fixed $i$, we have to determine which vertices of $V_i$ belong to $S^\star$; once more, since we do not know these vertices, we simply *guess* them; there are at most $n^{\lfloor k/p \rfloor}$ possible subsets to try because we assumed that $V_i$ contains at most $\lfloor k/p \rfloor$ vertices of $S^\star$. For each such subset $T \subseteq V_i$, we consider the subgraph $G' = (G - V_i) \cup T$. The treewidth of this subgraph is at most $c_H p + \lfloor k/p \rfloor$, and hence, we can find an optimal solution in $G'$ in time $\mathcal{O}(2^{\mathcal{O}(c_H p + k/p)} n^{\mathcal{O}(1)})$ and we are done.

The algorithm's total running time is $\mathcal{O}(2^{\mathcal{O}(c_H p + k/p + k \log n/p)} p n^{\mathcal{O}(1)})$. This expression is minimized for $p = \left\lfloor \sqrt{k \log n / c_H} \right\rfloor$, and so the algorithm runs in time $\mathcal{O}(2^{\mathcal{O}(\sqrt{c_H k \log n})} n^{\mathcal{O}(1)})$. Since for any fixed $\delta' > 0$ we have that $2^{\sqrt{\log n}} = o(n^{\delta'})$, we can choose $\delta'$ in such a way that for any given fixed $\delta > 0$, the running time is $o(n^{\mathcal{O}(1)+\delta\sqrt{k}})$.

On the other hand, for any $\alpha \geq 1$, if $c_H k \leq \alpha^2 \log n$, we have $\sqrt{c_H k \log n} \leq \alpha \log n$; and if $c_H k > \alpha^2 \log n$, we have $\sqrt{c_H k \log n} < c_H k/\alpha$. Hence, the running time is bounded by $\mathcal{O}(2^{\mathcal{O}(c_H k/\alpha)} + n^{\mathcal{O}(\alpha)})$. By choosing $\alpha = \Theta(c_H/\ln(1+\epsilon)) = \Theta(c_H/\epsilon)$, we obtain our result. □

Using the result of Dorn et al. [28] that the following problems are in EPT on (some) $H$-minor-free graphs when parameterized by treewidth, we immediately obtain:

**Corollary 12.** *For any graph $H$, the problem* DIRECTED $k$-PATH *are in* SUBEPT$^+$ *when restricted to $H$-minor-free graphs; the same is true for $k$-*STEINER TREE *at least on bounded-genus graphs*[3].

The two problems mentioned above are prominent problems that were not known to admit FPT-algorithms with running time better than $\mathcal{O}(2^k n^{\mathcal{O}(1)})$ before, even on planar graphs. Besides improving on the best known FPT-algorithms for these problems, our result shows that it is very likely that they indeed admit subexponential FPT-algorithms.

### 4.2 Guess and Conquer for Domination, Covering, and More

We introduced our technique for the class of $k$-{D,R,L,E,I}SP$(\pi)$ problems, where we are looking for a subgraph with a certain property. Whereas many problems can be formulated as an instance of this generic problem class, some others like $k$-VERTEX COVER, $k$-DOMINATING SET, or $k$-LEAF TREE and variants can not. We capture another class of problems by the following theorem.

**Theorem 13.** *Let $\Pi$ be a problem that takes as input a graph $G$ and outputs a set $S \subseteq V$ of vertices, and let $k$-$\Pi$ its parameterization by $|S|$. Suppose that*

*(i) on graphs of treewidth $t$, $\Pi$ can be solved in time $\mathcal{O}(2^{\mathcal{O}(t)} n^{\mathcal{O}(1)})$; and*
*(ii) if for an edge $e \in E(G)$ it is known that some solution of $S$ excludes both endpoints of $e$ then $\Pi$ can be reduced to finding a solution in $G - e$.*

*Then for any graph $G$ excluding a fixed minor $H$ there exists an algorithm $\mathcal{A}$ solving $k$-$\Pi$ on instance $(G, k)$ such that for any $\alpha \geq 1$, $0 < \epsilon \leq 1$, and fixed $\delta > 0$, algorithm $\mathcal{A}$ runs in time $\mathcal{O}_H(2^{\mathcal{O}_H(\sqrt{k \log n})} n^{\mathcal{O}(1)}) = \mathcal{O}_H(2^{\mathcal{O}_H(k/\alpha)} + n^{\mathcal{O}(\alpha)}) = \mathcal{O}_H((1+\epsilon)^k + n^{\mathcal{O}_H(1/\epsilon)}) = o(n^{\mathcal{O}(1)+\delta\sqrt{k}})$. In particular, $k$-$\Pi$ belongs to SUBEPT$^+$.*

*Proof.* Let $p$ be a fixed integer and let $E_1, \ldots, E_p$ be the edge partition obtained by Theorem 7, so that the graph induced by any $p - 1$ of the parts has treewidth at most $c_H p$ and furthermore, each vertex appears in at most $d_H$ of the parts. Let $S^\star$ be an optimal solution to $\mathcal{P}$ having at most $k$ vertices; then the total number of appearances of vertices in $S^\star$ in the parts $E_1, \ldots, E_p$ is bounded by $d_H k$, where $d_H$ is the constant from Theorem 7. It follows that there exists an $i^\star \in \{1, \ldots, p\}$ such that the graph induced by $E_{i^\star}$ contains at most $\lfloor d_H k/p \rfloor$ vertices of $S^\star$. We *guess* the value of $i^\star$ and the set of vertices $T^\star := S^\star \cap E_{i^\star}$ by trying all $pn^{\lfloor d_H k/p \rfloor}$ possibilities.

Because of assumption (ii), we can delete all the edges in $E_{i^\star}$ that do not have an endpoint in $T^\star$. The graph $G - E_{i^\star}$ has treewidth at most $c_H p$ and by adding the vertices of $T^\star$ to every bag in such a tree decomposition, the width becomes at most $c_H p + \lfloor d_H k/p \rfloor$. By choosing $p := \lfloor \sqrt{k \log n} \rfloor$ and repeating the analysis in the proof of Theorem 11, we obtain our result. □

The VERTEX COVER problem satisfies property (ii) above because if for an edge $e$, we know that both endpoints do not belong to the solution, then we can reject, since $e$ is not covered. For DOMINATING SET, such an edge is simply irrelevant, even for the connected version. That CONNECTED DOMINATING SET fulfills property (i) was shown by Dorn et al. [28] (see footnote on previous page). Hence, we have

**Corollary 14.** ({CONNECTED, INDEPENDENT}) $k$-DOMINATING SET *and* ({CONNECTED, INDEPENDENT}) $k$-VERTEX COVER *(at least) in bounded-genus graphs belong to the class* SUBEPT$^+$.

---

[3] In [28] it is claimed that STEINER TREE is in EPT on $H$-minor-free graphs when parameterized by treewidth; however, I know by private communication that at this time, a proof actually exists only up to bounded-genus graphs. The same is true for CONNECTED DOMINATING SET.

Still, Theorems 11 and 13 do not capture all problems to which the basic idea of our technique applies; for example, a modification of the proof of Theorem 13 shows that the technique also works for the undirected $k$-LEAF TREE problem. But since this problem is known to be in SUBEPT by the theory of bidimensionality, we refrain from presenting the details. It would be interesting to see if (a modification of) our technique can be used to solve the directed version of this problem.

Another interesting problem is $k$-BOUNDED DEGREE DELETION$(d)$, or $k$-BDD$(d)$ for short, where we want to delete a set of at most $k$ vertices so that the remaining graph has degree at most $d$. Note that $k$-BDD$(0)$ is equivalent to $k$-VERTEX COVER. Whereas we can not ignore edges that are known not to have endpoints in the solution, we can delete such edges and store at each vertex, the maximum allowed degree that remains; this information can then be incorporated in the dynamic programming on the bounded treewidth graph. The problem has a linear kernel and is thus in SUBEPT on $H$-minor-free graphs but for the case where we seek a connected solution, we obtain that $k$-BDD$(d)$ belongs to SUBEPT$^+$ only by applying our technique.

### 4.3 Further Analysis and Relation to Kernels

The analysis in the proof of Theorem 11 reveals that if $k = \mathcal{O}(\log n)$, then our algorithm runs in polynomial time; on the other hand, if $k = \omega(\log n)$, i.e. if $k$ is known to be at least $\Omega(\iota(n) \log n)$ for any computable, non-decreasing and unbounded function $\iota : \mathbb{N} \to \mathbb{N}$, then we have a SUBEPT algorithm with time complexity $\mathcal{O}(2^{\mathcal{O}(k/\iota(k))} + n^{\mathcal{O}(1)})$. But the condition $k = \omega(\log n)$ is nothing else but asking for a *subexponential kernel*; hence, we have:

**Corollary 15.** *Let $k$-Π be a parameterized problem that can be solved in time $\mathcal{O}(2^{\mathcal{O}(\sqrt{k \log n})} n^{\mathcal{O}(1)})$ and admits a minor-monotone subexponential kernel, i.e. can be reduced in polynomial time to an equivalent instance of size at most $2^{\mathcal{O}(k/\iota(k))}$ via edge-contractions and deletions for some computable, non-decreasing and unbounded function $\iota : \mathbb{N} \to \mathbb{N}$. Then $k$-Π belongs to SUBEPT. In particular, if $k$-Π admits a polynomial kernel then it can be solved in SUBEPT-time $\mathcal{O}(2^{\mathcal{O}(\sqrt{k \log k})} n^{\mathcal{O}(1)})$.*

Any parameterized problem that can be solved in time $\mathcal{O}(f(k) n^{\mathcal{O}(1)})$ admits a kernel of size $f(k)$ [23]. It follows that all problems in SUBEPT also have a subexponential kernel. Our corollary above shows the reverse direction of this observation for the problems that admit our technique on $H$-minor-free graphs; for these problems, we obtain that a subexponential FPT algorithm exists if and only if a minor-monotone subexponential kernel can be constructed.

Note that in the statement of Theorems 11 and 13, the parameter $\alpha$ is not required to be fixed; it can be any non-decreasing function from $\mathbb{N}$ to $\mathbb{N}$. Hence, the running time of the algorithm $\mathcal{A}$ that is obtained by these theorems cannot only be bounded subexponentially in $k$ but instead, slightly super-polynomially in $n$. For example, by choosing $\alpha = \log \log n$, we obtain a bound of $\mathcal{O}(2^{\mathcal{O}(k/\log \log k)} + n^{\mathcal{O}(\log \log n)})$.

## 5 Algorithms on Odd-Minor-Free Graphs

In [20], Demaine et al. prove a structural decomposition theorem for odd-minor-free graphs that is very similar to the RS-decomposition theorem for $H$-minor-free graphs [10]. They also present an algorithm running in time $n^{\mathcal{O}_H(1)}$ to compute such a decomposition. However, upon inspecting their proof, we obtain the following simpler intermediate result that turns out to be more useful for algorithmic purposes when combined with known results on $H$-minor-free graphs; in particular, it can be used to obtain FPT-versions of various algorithms when combined with our results from Section 3. Let $\mathscr{B}(\mu)$ denote the class of all bipartite graphs augmented by at most $\mu$ additional vertices called apices. We have

**Theorem 16 (adapted from [20]).** *Let $G$ be a given odd-$H$-minor-free graph. There exists a fixed graph $H'$ depending only on $H$ and an explicit uniform algorithm that computes a tree decomposition with adhesion at most $\kappa$ of $G$ that is strongly over the union of $\mathscr{B}(\mu)$ and the class of $H'$-minor-free graphs, where $\mu$ and $\kappa$ are computable functions depending only on $H$. Furthermore, we have the following properties:*

 (i) *the $H'$-minor-free graphs appear only in the leaves of the tree decomposition;*
 (ii) *if $B_2$ is a bag that is a child of the bag $B_1$ in the tree decomposition and $\overline{B}_1$ consists of a bipartite graph $W$ together with at most $\mu$ apices then $|B_2 \cap V(W)| \leq 1$;*
 (iii) *the at most $\mu$ apices of each bag are also computed;*
 (iv) *the algorithm runs in time $\mathcal{O}_H(n^4)$.*

*Proof.* The decomposition algorithm of Demaine et al. [20, Theorem 4.1] basically works as follows: if the given graph contains a certain fixed bipartite graph $H'$ as a minor, find a bipartite graph with some apices, create one bag out of it, and recurse on the components of the remaining graph; otherwise the graph excludes $H'$ as a minor and the RS-decomposition can be applied. Now instead of applying the RS-decomposition at this step (which is Step (7) of the algorithm in [20]), we simply create a bag containing the current subgraph and connect it as a leaf to the tree decomposition. All other properties follow directly from [20, Theorem 4.1] and its proof and analysis. □

Together with Theorem 3 we obtain the following corollary:

**Corollary 17.** *There is an explicit uniform algorithm that, given an odd-$H$-minor-free graph $G$, computes a tree decomposition $(T, \mathcal{B})$ of $G$ with adhesion at most $\kappa$ in time $\mathcal{O}_H(n^{\mathcal{O}(1)})$ such that for every bag $B \in \mathcal{B}$, we have either*

*(i) the companion of $B$ is in $\mathcal{L}(\lambda, \mu)$; or*
*(ii) the closure of $B$ is in $\mathscr{B}(\mu)$,*

*where $\lambda$, $\mu$, and $\kappa$ are computable functions depending only on $|H|$. The $\mu$ apices of (the companion of) each bag in $\mathcal{B}$ can be computed in the same time bound. Moreover, if $B_1, B_2 \in \mathcal{B}$ and $B_2$ is a child of $B_1$ in the tree decomposition and $\overline{B}_1$ consists of a bipartite graph $W$ together with at most $\mu$ apices then $|B_2 \cap V(W)| \leq 1$.*

*Proof.* We first apply Theorem 16 to obtain a tree decomposition $(T, \mathcal{B})$ as described. Then we consider each leaf that contains an $H'$-minor-free subgraph $G'$ of $G$ and apply Theorem 3 to it to obtain a tree decomposition $(T_{G'}, \mathcal{B}_{G'})$ that is weakly over $\mathcal{L}(\lambda, \mu')$; afterwards, we add the intersection of $V(G)'$ with its parent bag in $(T, \mathcal{B})$ to each bag of $\mathcal{B}_{G'}$ and replace the leaf of $(T, \mathcal{B})$ containing $G'$ with this finer tree decomposition. This way, we make sure that the number of apices in every bag of the global tree decomposition is the maximum of the value obtained from Theorem 16 and $\mu' + \kappa$ and let $\mu$ be this maximum. □

The proof of the following theorem is analogous to the proof of Theorem 7; only note that the vertex set of a graph from $\mathscr{B}(\mu)$ naturally has a partition into 2 parts of bounded treewidth: just take each part of the bipartition together with some apices (see also the proof of [20, Theorem 1.1]).

**Theorem 18.** *For any fixed graph $H$ there is a constant $c_H$ such that for every odd-$H$-minor-free graph $G$, the vertices of $G$ can be partitioned into 2 parts such that each of the parts induces a graph of treewidth at most $c_H$. Furthermore, such a partition can be found in explicit uniform FPT-time, i.e. $\mathcal{O}_H(n^{\mathcal{O}(1)})$.*

This is the best possible analog to the Baker-style decomposition of Theorem 6 for odd-minor-free graphs since these graph classes include all bipartite graphs; and complete bipartite graphs can not be partitioned into more than 2 parts of bounded treewidth. A direct corollary is the following:

**Corollary 19.** *There exists a 2-approximation algorithm for COLORING an odd-$H$-minor-free graph in time $\mathcal{O}_H(n^{\mathcal{O}(1)})$.*

Also, 2-approximations with the same FPT-running time for various other problems, such as many of the ones mentioned in Section 3.4, can be obtained. See [7] and [20] for more details.

## 5.1 PTASes on Odd-Minor-Free Graphs

Grohe [4] showed that various problems admit a PTAS on $H$-minor-free graphs. Most of these PTASes *can not* be generalized to odd-minor-free graphs as they would imply corresponding PTASes on bipartite or even general graphs for APX-hard problems. However, Demaine et al. ask several times in [20] whether the PTASes for VERTEX COVER and INDEPENDENT SET can be generalized to odd-minor-free graphs; this seems plausible since these two problems can be solved in polynomial time on bipartite graphs. Indeed, we are able to answer this question affirmatively in this section. To this end, we define the *take-or-leave* version of these problems as follows: every vertex of the graph is associated with two numbers $w^+$ and $w^-$; if a vertex is chosen to be in the solution, i.e. in the vertex cover or independent set, it contributes a value of $w^+$ to the objective function; if it is not included in the solution, it contributes $w^-$ to the objective function (the usual unweighted variants are then special cases of the take-or-leave version where $w^+ = 1$ and $w^- = 0$ for every vertex).

**Lemma 20.** *The take-or-leave versions of VERTEX COVER and INDEPENDENT SET can be solved in polynomial time on bipartite graphs.*

*Proof.* Just note that the matrices used in the standard linear programming formulations of the unweighted version of these problems are totally unimodular for bipartite graphs, and hence all the corners of the corresponding polyhedra are integral. But the only thing that changes now is the objective function; in particular, the polyhedron is still the same and integral. Hence, we can find a solution in polynomial time just by solving these linear programs. □

**Theorem 21.** *There exists a PTAS for* VERTEX COVER *and* INDEPENDENT SET *in odd-$H$-minor-free graphs running in time $\mathcal{O}_{H,\epsilon}(n^{\mathcal{O}(1)})$.*

*Proof.* We prove the theorem for VERTEX COVER; the case of INDEPENDENT SET is analogous. Let $G$ be a given odd-$H$-minor-free graph. We first compute a tree decomposition $(T, \mathcal{B})$ of $G$ as specified by Theorem 16 and root it at some bag containing a graph from $\mathscr{B}(\mu)$; if such a bag does not exist, the graph is actually $H'$-minor-free and we obtain our result by Corollary 9. For every bag $B \in \mathcal{B}$ we define the *subproblem at $B$* to be the considered problem on the subgraph of $G$ that is induced by $B$ and all of its descendants in $T$. We perform dynamic programming from the leaves of the tree decomposition to the root and store at each bag $B$ the following information: for every subset $U$ of the apices $A$ of $B$, we compute a solution for the subproblem at $B$ that must contain $U$, must not contain $A - U$, and is within a factor of $(1 + \epsilon)$ of the optimal solution with these properties; we let $\nu(B, U)$ be the value of such a solution minus $|U|$ and store it in a table; if a solution with these properties does not exist, we store $\bot$ to denote this fact. Here we assume w.l.o.g. that we have nice apex sets that include the intersection of the current bag with its parent bag; in fact, for leaves of the tree that contain an $H'$-minor-free graph, the apex set is defined to be this intersection. For such leaves, we can compute the values of the dynamic programming table by invoking $2^\kappa$ times the PTAS from Corollary 9 – once for each subset $U \subseteq A$.

Now suppose we are at a bag $B$ with children $B_1, \ldots, B_t$ ($t \geq 0$) and $B$ contains a graph from $\mathscr{B}(\mu)$. Suppose $B$ contains a bipartite graph $W$ together with apices $A$. For each fixed selection $U \subseteq A$ of the apices, we have to compute a near optimal solution $\nu(B, U)$. For each child $B_i$, define $Y_i = B_i \cap B$ and for each set $X \subseteq Y_i$, let $\nu^\star(B_i, X)$ be the value of a $(1 + \epsilon)$-approximate solution for the subproblem at $B_i$ that contains $X$ but not $Y_i - X$, minus $|X|$ (or $\bot$ if such a solution does not exist). Note that the value of $\nu^\star(B_i, X)$ can be looked up in the dynamic programming table of $B_i$ by taking the minimum over all $\nu(B_i, U') + |U'| - |X|$ with $U' \cap Y_i = X$. These values can be precomputed and stored.

Note that the status of every vertex in $Y_i$, i.e. whether or not it should be in the solution, is completely specified by the choice of $U$ except for at most one vertex $v \in Y_i \cap V(W)$. For each vertex $v \in V(W)$, define its *taking* weight $w^+$ as $1 + \sum \nu^\star(B_i, \{v\} \cup (U \cap Y_i))$ and define its *leaving* weight $w^-$ as $\sum \nu^\star(B_i, U \cap Y_i)$ where the sums go over all children $B_i$ that contain $v$. We solve the take-or-leave version of the problem with these weights on $W$ using Lemma 20 and store it as the solution for $\nu(B, U)$ (or $\bot$ if no such solution exists). We return the solution of minimum value stored at the root of the tree decomposition. This finishes the description of the algorithm.

The correctness is immediate at the leaves of the dynamic program. For a non-leaf bag $B \in \mathcal{B}$ with bipartite graph $W$ and apices $A$ and a given selection $U \subseteq A$ of its apices, let $S^\star$ be an optimal solution for the subproblem at $B$ that includes $U$ but not $A - U$, and $S$ be the solution corresponding to $\nu(B, U)$ as computed by our algorithm. For a set $X \subseteq V(W)$, let $\mathrm{OPT}_X$ denote the optimal solution value for the subproblem at $B$ given that $U \cup X$ must be in the solution and $B - U - X$ must not be in the solution; let $\mathrm{TOL}_X$ denote the objective function value of the take-or-leave problem defined at $B$ if $X$ is taken as the solution, plus $|U|$. By our construction and the induction hypothesis of the dynamic program, for any set $X \subseteq V(W)$, we have $\mathrm{TOL}_X \leq (1 + \epsilon)\mathrm{OPT}_X$. Hence, we obtain $|S| = \mathrm{TOL}_{S \cap V(W)} \leq \mathrm{TOL}_{S^\star \cap V(W)} \leq (1 + \epsilon)\mathrm{OPT}_{S^\star \cap V(W)} = (1 + \epsilon)|S^\star|$. □

Note that Theorem 21 holds also for the vertex-weighted versions of these problems; the proof is analogous.

### 5.2 Subexponential FPT for Odd-Minor-Free Graphs

Another question that is asked by Demaine et al. [20] is whether VERTEX COVER and INDEPENDENT SET admit subexponential FPT-algorithms on odd-minor-free graphs. As in the case of the PTASes, these are basically the only problems for which this seems possible as such algorithms for most other prominent problems would contradict hardness results in parameterized complexity. Indeed, we can obtain subexponential parameterized algorithms for these problems in a similar way as the PTASes above. First, let us state the following known result[4].

---

[4] I would like to thank Fedor Fomin for a helpful discussion on this matter.

**Lemma 22 (partly taken from [35,14]).**

*(i) There exists an algorithm that given an $H$-minor-free graph $G$ and an integer $k$ decides if $G$ contains a vertex cover of size at most $k$ and in this case, returns a minimum vertex cover of $G$ in time $\mathcal{O}(2^{\mathcal{O}_H(\sqrt{k})} n^{\mathcal{O}(1)})$.*

*(ii) There is an algorithm with the same time complexity that decides if $G$ contains an independent set of size at least $k$ and if this is* not *the case, returns an independent set of maximum size in $G$.*

*Proof.* The algorithm for VERTEX COVER follows directly from the bidimensionality theory [35,14].

As for INDEPENDENT SET, it is a well-known fact that $H$-minor-free graphs have bounded average degree [36] and hence, an independent set of size $\Omega_H(n)$. So, all we have to do is to count the number of vertices of $G$; if this is at least $c_H k$, for a suitable constant $c_H$, the answer is "yes". Otherwise, the size of the graph is bounded by $c_H k$ and hence has treewidth at most $\mathcal{O}_H(\sqrt{k})$ [37]; an optimal independent set can be found by standard dynamic programming. □

**Theorem 23.** *There exists an algorithm that, given an odd-$H$-minor-free graph $G$ and an integer $k$, runs in time $\mathcal{O}(2^{\mathcal{O}_H(\sqrt{k})} n^{\mathcal{O}(1)})$ and*

*(i) decides if $G$ contains a vertex cover of size at most $k$ and in this case, returns a minimum vertex cover of $G$; and*

*(ii) decides if $G$ contains an independent set of size at least $k$ and if this is* not *the case, returns an independent set of maximum size in $G$.*

*Proof.* The proof is basically analogous to the proof of Theorem 21 above. We start by computing a tree decomposition as given by Theorem 16. For each bag $B$ of and each selection $U \subseteq A$ of the apices of the bag, we compute a value $\nu(B, U)$, specifying if the subproblem at $B$ contains a vertex cover of size at most $k$ that includes $U$ but not $A - U$ and if so, the size of the minimum vertex cover with these properties; if not, we store $\bot$ to indicate a negative answer. If all entries for a bag $B$ turn out to be $\bot$, the answer to problem is "no" and we can terminate.

The entries for the leaves that contain an $H'$-minor-free graph can be computed using Lemma 22. For a (leaf or non-leaf) bag $B$ that contains a graph from $\mathcal{B}(\mu)$ we construct a take-or-leave version of VERTEX COVER analogously to what we did in the proof of Theorem 21 and solve it in polynomial time. The correctness follows as in Theorem 21 and the running time is dominated by the running time of the algorithm on the $H'$-minor-free leaves and hence is subexponential FPT in $k$.

The case for INDEPENDENT SET is analogous only that if at any point the answer to some subproblem is "yes" we may return this answer and terminate; otherwise, we proceed as above. □

## 6 Conclusion

We significantly accelerated one of the main tools in PTAS design – namely, Baker's decomposition – on all proper minor-closed graph classes, thereby obtaining the improvement for all the applications of this tool. We showed similar results for odd-minor-free graphs and obtained the first PTAS and subexponential FPT-algorithms for VERTEX COVER and INDEPENDENT SET on these graph classes. Based on Baker's approach, we also introduced the new technique of Guess & Conquer for designing (nearly) subexponential FPT-algorithms on these graph classes. We improved the best known FPT-algorithms for STEINER TREE and DIRECTED $k$-PATH that were previously only known to be in EPT even on planar / apex-minor-free graphs. We actually conjecture that these problems are in SUBEPT on $H$-minor-free graphs and repeat this as an important open question for future work.

**Acknowledgment**. I would like to thank Holger Dell, Martin Grohe, and Matthias Mnich for helpful discussions and comments on earlier versions of this work.